\documentclass{acmtrans2e}
\usepackage{url,graphicx}

\firstfoot{ACM Transactions on Information and System Security, Vol.~5, No.~3,
August 2002, Pages 262--289.}

\runningfoot{ACM Transactions on Information and System Security, Vol.~5,
No.~3, August 2002.}

\title{Information Leakage from Optical Emanations}

\author{JOE LOUGHRY \\ Lockheed Martin Space Systems
\and DAVID A.\ UMPHRESS \\ Auburn University}

\begin{abstract}

A previously unknown form of compromising emanations has been discovered. LED 
status indicators on data communication equipment, under certain conditions,
are shown to carry a modulated optical signal that is significantly correlated
with information being processed by the device. Physical access is not
required; the attacker gains access to all data going through the device,
including plaintext in the case of data encryption systems. Experiments show
that it is possible to intercept data under realistic conditions at a
considerable distance.  Many different sorts of devices, including modems and
Internet Protocol routers, were found to be vulnerable.  A taxonomy of
compromising optical emanations is developed, and design changes are
described that will successfully block this kind of ``Optical \textsc{Tempest}''
attack.

\end{abstract}







\begin{document}
\begin{bottomstuff}

Much of this work was done while the first author was a graduate student
in the Department of Computer Science and Software Engineering at Seattle
University.

Authors' addresses: Joe Loughry, Lockheed Martin Space Systems,
Dept.\ 3740, Mail Stop X3741,
P.O.\ Box 179, Denver, Colorado 80201 USA, email
\url{joe.loughry@applied-math.org};
David A.\ Umphress, Auburn University, Department of Computer Science and
Software Engineering, 215 Dunstan Hall, Auburn University AL 36849 USA,
email \url{umphress@eng.auburn.edu}.

\permission
\end{bottomstuff}

\maketitle

\section{Introduction}

Can optical radiation emitted from computer LED (light emitting diode)
status indicators compromise information security?  Data communication
equipment, and even data encryption devices, 
sometimes emit modulated optical signals that carry enough information 
for an eavesdropper to reproduce the entire data stream being processed 
by a device.  It requires little 
apparatus, can be done at a considerable distance, and is completely 
undetectable.  In effect, LED indicators act as little free-space 
optical data transmitters, like fiber optics but without the fiber.

\begin{figure}
\centering
\includegraphics[height=2in]{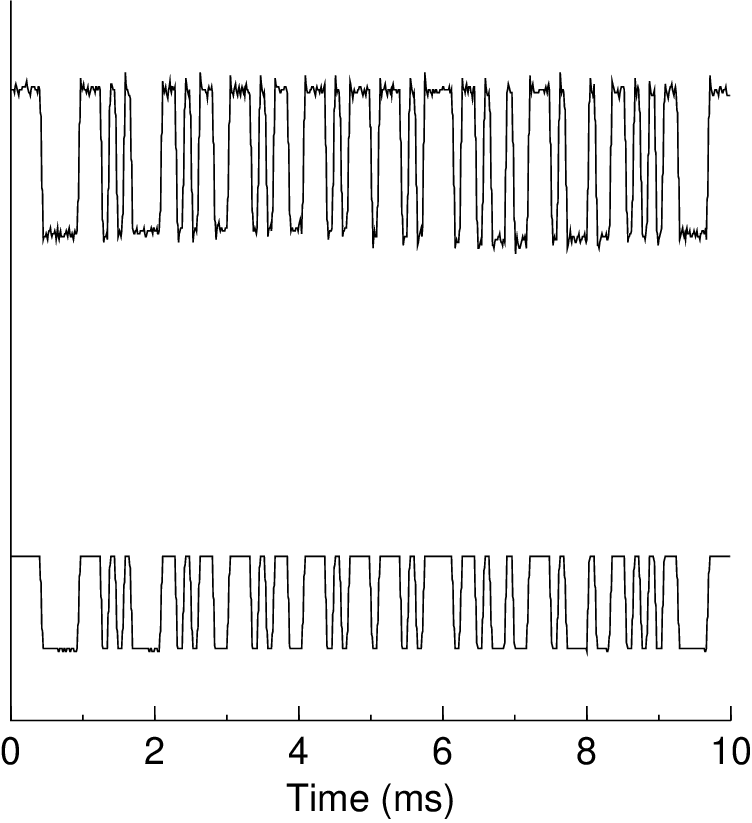}
\caption{Compromising optical emanations.  The lower trace shows the
$\pm15$V EIA/TIA-232-E input signal (at 9600 bits/s); the upper
trace shows optical emanations intercepted 5~m from the device.}
\label{example_of_optical_emanations_figure}
\end{figure}

Experiments conducted on a wide variety of devices show evidence of 
exploitable compromising emanations in 36\% of devices tested.  With 
inexpensive apparatus, we show it is possible to intercept and read data 
under realistic conditions from at least across the street.  In Figure 
\ref{example_of_optical_emanations_figure}, the lower trace shows the
$\pm15$V EIA/TIA-232-E waveform of a serial data signal at
$9600\ \mathrm{b}/\mathrm{s}$. 
The upper trace shows modulated optical radiation intercepted
$5\ \mathrm{m}$ from the device.  A high correlation is evident.

We have successfully recovered error-free data at
speeds up to $56\ \mathrm{kb}/\mathrm{s}$; the physical principles
involved ought to
continue to work up to about 10 Mbits/s.  Protecting against the 
threat is relatively straightforward, but may require design
changes to vulnerable equipment.

\subsection{Paper Organization}

The first part of this paper reviews the idea of compromising 
emanations, and gives an overview of what information is to be found in 
the literature.  Next comes a technical explanation of why compromising 
optical emanations exist, together with some of their properties.  A series of 
experiments is then described, along with results that were found.  
Finally, some possible countermeasures are discussed, along with 
directions for future work. Related work on active attacks using optical 
emanations is presented in the appendices.

\section{EMSEC, TEMPEST, and Compromising Emanations}

\begin{quotation}
{\bf Compromising Emanations} \cite{ncsc_tg_004}: ``Unintentional
data-related or intelligence-bearing signals that, if 
intercepted and analyzed, disclose the information transmi[tted],
received, handled, or otherwise processed by any information processing 
equipment. See {\it TEMPEST}.''
\end{quotation}

Thorough discussion of compromising emanations and EMSEC (emissions
security) in the open literature is limited.  The information that is 
available tends to exhibit a strong bias toward radio frequency 
(RF) emanations from computers and video displays.  Because of the high 
cost of equipment and the difficulty of intercepting and exploiting RF 
emanations, reports of successful attacks against emanations 
have been limited primarily to high-value sources of information such 
as military targets and cryptologic systems.  A significant problem is
that much important information on compromising emanations is
classified \cite{russell_and_gangemi}, although some documents have
recently been declassified \cite{NACSIM_5000,NSTISSAM,shielded_enclosures}.

\subsection{Related Work}

The ability to compromise signals emanating from computers has been known
for some time.  For instance, Smulders \citeyear{smulders} found
RF emanations in unshielded or poorly shielded serial cables, and
van~Eck \citeyear{van_Eck} showed that cathode-ray tube video
displays can be read at a distance by intercepting and analyzing
their RF emanations.  Others have noted RF compromise, including
more contemporary research showing ways to hide
information in signals emitted by video devices as well as specialized
fonts that minimize compromising RF emanations \cite{kuhn_and_anderson}.
Wright \citeyear{wright} described, anecdotally, the discovery of
electrically conducted compromising emanations from cipher machines as
early as 1960.  For an excellent overview of the current state
of emanations security research, the interested reader is referred to 
the book by Anderson \citeyear{anderson_security_engineering} and a
related paper by Kuhn and Anderson \citeyear{kuhn_and_anderson}.

Until recently, little mention of signals in the optical spectrum was
found in the literature.  Kuhn \citeyear{markus} has demonstrated
remotely reading CRT displays from diffuse optical emanations, without
requiring line-of-sight access. Other related topics include security
of fiber optics
\cite{hodara_1991,exfo_data_sheet} and optical communications
\cite{wilkins_1641}.  Social engineering attacks such as ``shoulder
surfing'' and visual surveillance of video displays are well covered
in \cite{fites_and_kratz}.  Free-space optical data links are prone
to interception, and  for this reason wireless data links (both laser
and RF) are typically encrypted \cite{lathrop}.  But with the exception
of a work of fiction, in which one character uses the LEDs on a computer
keyboard\footnote{See also Appendix~\ref{appendix:covert_channel}.}
to send information in Morse code \cite{cryptonomicon}, and inferences
from redacted sections of partially declassified
documents \cite{NACSIM_5000}, a thorough search of the literature
revealed no direct mention of the risk of interception of data from
optical emanations of LED status indicators.

\section{Compromising Optical Emanations}

\begin{quote}
\emph{``The [IBM] 360 had walls of lights; in fact, the Model 75 had so
many that the early serial number machines would blow the console power 
supply if the `Lamp Test' button was pressed.''} \cite{morris}
\end{quote}

\subsection{Light-Emitting Diodes}

Light-emitting diodes are cheap, reliable, bright, and ubiquitous.  They 
are used in nearly every kind of electronics, anywhere a bright, 
easy-to-see indicator is needed.  They are especially common in
data communication equipment.  Every year, some 
20--30 billion LEDs are sold \cite{led_shipments}.

LEDs are very fast; that is, they exhibit a quick response to 
changes in the applied drive voltage (tens of nanoseconds).  In fact,
common visible LEDs
are fast enough that a close cousin is used as a transmitter in fiber
optic data links at speeds in excess of 100~Mbits/s
\cite{hp_fiber_optic_data_sheet}.

Although fast response time is oftentimes a desirable quality in a 
display, LEDs are fast enough to follow the individual bit transitions 
of a serial data transmission.  Herein lies the problem: if certain LED 
indicators are visible to an attacker, even from a long distance away,
it becomes possible for that person to read all of the data going through the 
device.

One of the advantages of LED displays is that they can be read from 
across a room.  The disadvantage may be that they can be read from across 
the street.

\subsection{Rationale for the Existence of Compromising Optical 
Emanations}

\begin{figure}
\centering
\includegraphics[height=2in]{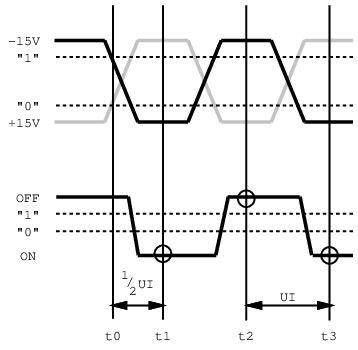}
\caption{EIA/TIA-232-E serial data waveform and typical LED response.}
\label{idealized_led_response_figure}
\end{figure}

The brightness of LED displays would not be a problem if it were not for 
the way they interact with serial data transmissions.  Consider the
idealized EIA/TIA-232-E waveform and associated LED response curve depicted in
Figure \ref{idealized_led_response_figure}.  The upper waveform shows
the EIA/TIA-232-E serial data signal; the lower waveform illustrates
the optical output of an LED indicator monitoring that signal.  As long
as the rise time of
the LED is less than $\frac{1}{2}$ of the unit interval $t_\mathrm{UI}$,
the LED will accurately enough mirror the EIA/TIA-232-E data signal at the
critical points shown by the small circles in the diagram to enable recovery
of the original data.

\begin{figure}
\centering
\includegraphics[height=2in]{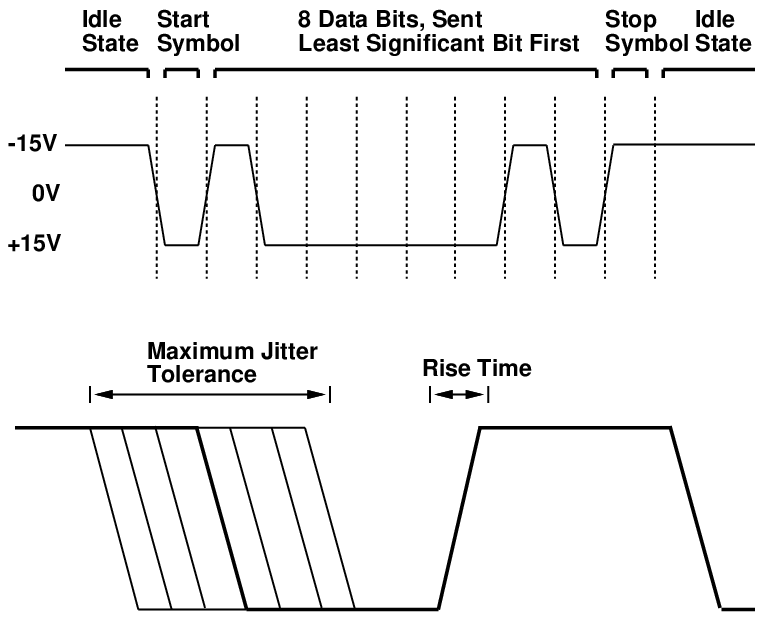}
\caption{EIA/TIA-232-E serial data waveform and maximum jitter
tolerance from TIA/EIA-404-B.}
\label{jitter_tolerance_figure}
\end{figure}

The EIA/TIA-232-E standard (formerly known as RS-232) defines a
bit-serial format using bipolar encoding and non-return-to-zero--level 
(NRZ--L) signaling \cite{eia_tia_232_e}.  As illustrated in Figure
\ref{jitter_tolerance_figure}, bits are
transmitted asynchronously, with framing 
bits embedded in the serial data stream for synchronization 
between sender and receiver.  During periods when no data are being 
transmitted, the transmitter remains in the logical ``1'' state.  The 
start of a new symbol is indicated by a momentary excursion to the 
logical ``0'' state for one unit interval, called the {\it start bit}.  
This is followed by a serial waveform consisting of a mutually
agreed-upon number of data bits, sent least significant bit first.  Following 
the last data bit, the transmitter returns to the logical ``1'' state 
for at least one unit interval, called the {\it stop bit}, in order to 
provide necessary contrast for the receiver to recognize the 
beginning of the next start bit.  (Another way of looking at this is that the
channel is required to return to the idle state for at least one unit interval
between characters.)

EIA/TIA-232-E uses bipolar encoding, with a negative voltage signifying 
logical ``1'' and a positive voltage used for logical ``0''
\cite{black}.  Usually, 
LEDs are wired to light up for a logical ``0'' so that they flicker when 
bits are transmitted, and remain dark when the channel is idle.  The 
fact that the original signal is bipolar is immaterial.  As long as 
the LED is fast enough to faithfully reproduce the timing of bit transitions, 
the optical output will contain all of the information in the 
original EIA/TIA-232-E signal.

LEDs cannot be connected directly to logic circuits, as they would draw too 
much power from the signal source.  For reasons of cost, however, the very same high-speed
gates (usually TTL or CMOS inverters) typically used to construct logic circuits
are also employed to power the LEDs \cite{ttl_cookbook}.
The result is a direct path allowing information to flow from the serial
data channel to the optical output of the LED.  Because the monitoring circuit
was not designed for the
purpose, the resulting optical signal may exhibit noise or other degradation,
but LEDs and their associated driver circuitry are generally more than fast
enough to reproduce a serial data signal at normal data rates.

\subsubsection{Characteristics of the Optical Signal}

NRZ--L signals are susceptible to 
noise, which is why other signaling methods, such as differential 
Manchester encoding, are most often used in long-distance digital 
communication systems.  To overcome the noise sensitivity of NRZ--L, 
additional redundancy is often introduced into the communication channel 
in the form of {\it channel encoding} \cite{proakis}.  Parity checks, 
cyclic redundancy checking (CRC), and other error detection and 
correction methods may be used to increase the reliability of the 
system.  But it should be noted that these features are also available 
to an eavesdropper, who may use them to overcome the effects of a poor 
optical signal.

As optical communication systems go, it must be recognized that LED
status displays are highly sub-optimal.  There are no beam-forming
optics on the transmitting LED.  The radiant flux available is
extremely limited.  Buffer circuits used to drive LED indicators, while 
more than fast enough for their intended purpose, are not optimized for 
high-speed data transmission in the way that special-purpose circuits 
used in fiber optic transmitters are.  Practical optical data 
communication systems use laser transmitters, sophisticated encoding 
schemes, and coherent detectors that greatly improve signal recovery 
under noisy conditions \cite{gagliardi}.  Our hypothetical eavesdropper 
would likely have to deal with off-axis aiming errors, high levels of 
optical background noise from artificial lighting, and lack of a
priori knowledge of the specific bit rate and word length used by the target.  
Nevertheless, our experiments show that with a sensitive detector and 
telescopic optics, it is possible for an eavesdropper to recover a noisy 
analog waveform closely approximating the original digital data stream.  
Once the received optical signal has been amplified, cleaned of noise, 
and fed to a USART (Universal Synchronous-Asynchronous
Receiver-Transmitter)---an inexpensive chip which serves as a ready-made
solution to the problem of decoding a noisy signal---the original data
stream is easily recovered.

\subsubsection{Insensitivity to the Modulation Scheme Employed}

High-speed modems employ a variety of complicated modulation schemes, including 
frequency, amplitude, and phase modulation to maximize available 
bandwidth on voice-grade telephone lines.  But this makes no
difference---it is the relatively simple NRZ--L waveform of the
EIA/TIA-232-E data signal that is modulated onto the LED.

\subsubsection{Nonsusceptibility of Other Light Sources}

Questions remain as to the susceptibility of non-LED sources to
interception of compromising optical emanations.  Liquid crystal
(LCD) displays, in particular, exhibit a relatively slow impulse
response, typically on the order of tens of milliseconds, making
these displays relatively poor sources of compromising optical
emanations, except at fairly low data rates.  Cathode ray tube
(CRT) displays, however, at the pixel level, are very fast, and
have been shown to be vulnerable even to non-line-of-sight optical
emanations \cite{markus}.


\begin{table}
\centering
\caption{Proposed classification system for optical emanations.}
\vspace{.1in}
\label{risk_table}
\begin{tabular}{|c|c|c|}
\hline
Type & Correlated to & Associated Risk Level
\rule{0in}{2.5ex} \\ [0.5ex]
\hline
Class~I & State of the device & Low \rule{0in}{2.5ex} \\
Class~II & Activity level of the device & Medium \\
Class~III & Content (data) & High \\
\hline
\end{tabular}
\end{table}

\subsection{Classification of Optical Emanations}

It is useful to consider a division of optical emanations into three 
broad classes according to the amount of information potentially carried 
to an adversary.  The proposed taxonomy is shown in Table \ref{risk_table}.
In the list that follows, LED indicators that exhibit
Class~$n$ behavior are called Class~$n$ indicators.

The classifications are:

\begin{itemize}

\item{\bf Class~I} indicators, which are unmodulated.  The
optical emanations put out by this type of display are constant, and
correlated with the {\it state} of a device
or communication channel.  Class~I indicators communicate at most one bit 
of information to an observer.  An example would be a power-on indicator.

\item {\bf Class~II} indicators are time-modulated, and correlated with
the {\it activity level} of a device or communication 
channel.  Class~II indicators provide an adversary with considerably 
more information than Class~I indicators do.  On face value, while the
content of the
data being processed by a device is not known, the fact that something is
being transmitted, and a  rough idea of where and how much, together
make possible {\it traffic analysis} of interesting targets.  Examples
of Class~II indicators include the {\tt Work Station Active} light on
an IBM 5394 Control Unit, activity indicators on Ethernet interfaces, and the
front-panel lights of a Cisco router.  It is important to note
that by affecting the activity level of a device, and hence modulating
the output of a Class~II indicator, it is possible for an attacker to
implement a covert timing channel.

\item {\bf Class~III} optical emanations are modulated optical
signals that are strongly 
correlated with the {\it content} of data being transmitted or received.  
If the correlation is sufficiently good, then from analysis of Class~III 
optical emanations it is possible to recover the original data stream. 
Examples of Class~III emanations are surprisingly common; the 
``Transmitted Data'' and ``Received Data'' indicators on modems
are usually Class~III.

\end{itemize}

Devices having at least one Class~II 
indicator, but no Class~III indicators, are called Class~II devices; 
any device having at least one Class~III indicator is a Class~III 
device.  Class~III devices are the most interesting.

Note that in both the Class~I and Class~II cases, the adversary gets no more 
information than the operator does; the indicator is being used in the 
manner for which it was intended, except that the eavesdropper is 
unauthorized, and reading the information at a distance.

Class~III devices may arise when the designer of a device inadvertently 
specified a Class~III indicator where a Class~II indicator was needed.
It is not clear whether 
there is {\it any} situation in which a Class~III indicator would be 
warranted, except in the case of an extremely low-speed communication 
channel, where individual bit transitions could be observed by eye and 
decoded.  In most cases the activity of a data communication channel 
occurs too fast for the human eye to follow.  In the real world, an 
oscilloscope is a much more useful tool than a Class~III indicator.

Potentially dangerous Class~III indicators can be converted to the safer 
and more useful Class~II type by the addition of a pulse stretching circuit,
as described in Section \ref{section:countermeasures}
on Countermeasures below.

\section{Eavesdropping Experiments}

Three series of experiments were run.  First, a survey was made of a 
large number of devices, looking for evidence of Class~III behavior.  
Then, long-range testing was done on a selection of devices, to prove the 
feasibility of interception under realistic conditions.  Finally, 
examination was made of the internals of several devices, in an attempt 
to understand why these emanations occur.

\subsection{Hypothesis}

The null hypothesis was stated as follows: ``It is not possible to recover
data from optical emanations.''  The null hypothesis was disproved
by experiment.
 
\subsection{Experimental Design and Methodology}

A total of 39 devices containing 164 unique LED indicators were 
identified for this study.  The devices selected for testing were chosen 
to represent a wide variety of information processing technology, 
including low-speed and high-speed communication devices, local-area 
network (LAN) and wide-area network (WAN) devices, PC and mainframe 
computers, mass storage devices, and peripherals.

Prior to commencement of measurements, radiometric readings were
taken on an optical bench of a standard 
red LED driven by a square wave signal.  These measurements were used to 
establish a baseline.  Following this step, each of the 164 LED 
indicators identified in the survey was examined for evidence of 
Class~III behavior.

Measurements were made of individual LED indicators by placing a hooded
detector in contact with each LED.  A dual-trace oscilloscope was used 
to observe the signal from the detector.  To visualize the corresponding 
data stream, a breakout box was inserted into the data path, with the 
original data displayed alongside the optical signal from the detector. 
 
The detector used was a high-speed, large-area silicon PIN
(Positive--Intrinsic--Negative) photodiode with an active area of
1~$\mathrm{mm}^2$.  
The responsivity of this detector is 0.45 A/W at a 
nominal wavelength of 830~nm, with a spectral response of 350--1100~nm.  
The photocurrent from the detector was amplified by a transimpedance 
photodiode amplifier operated in zero-bias mode.  Signals were 
observed with a 200~MHz digital oscilloscope, and captured for later
analysis.

The bandwidth of the 
photodiode amplifier is inversely proportional to its gain setting; at
a gain factor of $10^7$~V/A, the bandwidth of the
detector--amplifier system is only 10~KHz.  Therefore, for most
measurements, the 
amplifier was operated at a gain setting of $10^4$~V/A, yielding an 
overall detector--amplifier system bandwidth of 45~KHz, which was
marginal, but adequate.  For higher-speed 
measurements, the photodiode was connected directly to the input 
amplifier of the oscilloscope and operated in the quadrant IV 
(photovoltaic) region.  Limited sensitivity in this configuration 
is what necessitated placing the detector directly in contact with the 
LED.  However, the greatly increased bandwidth of the detector--amplifier 
system in this configuration allowed for examination of very high
speed devices for evidence of signals in the MHz range.

\subsubsection{Long-Range Testing}

Long-range optical eavesdropping experiments were conducted with a small 
number of representative devices.  The ANP Model 100 short-haul modem, 
Hayes Smartmodem OPTIMA 9600 and 14400, and a Practical Peripherals 
PM14400FXMT fax modem were all examined.

The same photodetector and amplifier system described in the previous 
section was used.  The detector was mounted at the 
focus of an optical system consisting of a 100~mm diameter, $f/2.5$ 
converging lens, an aperture stop, and a 650~nm optical bandpass filter, 
chosen to match the spectral output of a standard visible red 
LED \cite{agilent_led_data_sheet}.

The device under test was placed a measured distance away, and connected 
to an identical unit at the test station through a length of unshielded 
twisted pair cable.  The image from a single LED on the 
device under test was adjusted to completely cover the detector's active 
area.  Test transmissions were made to each device, and the
EIA/TIA-232-E waveform and resulting optical signals captured for
analysis.

\subsubsection{Experimental Methodology}

Three independent variables and one dependent variable were identified. 
The independent variables were: (1) the separation distance between the 
detector and the device under test, (2) the data transmission rate, and 
(3) ambient lighting conditions on the test range.  The dependent 
variable was the correlation between the received optical signal and the 
original EIA/TIA-232-E waveform captured at the same time.  The 
independent variables were varied according to a formal test matrix.  
Separation distance was varied from 5~m to 38~m (the maximum 
dimension of the laboratory) in increments of 5~m during the test.  
At each measured distance, test transmissions were made at data rates of 
300, 600, 1200, 2400, 4800, 9600, and 19~200~bits/s.

For simplicity, symbols in the optical signal were detected by observing
the signal's amplitude at one-half of the unit interval after the NRZ--L
transition.  Because this was a proof-of-concept experiment, actual
bit-error rates were not measured.  The optical
waveform from the detector amplifier was compared to the original
EIA/TIA-232-E signal waveform obtained from a breakout box 
inserted in the data path between the data generator and the device 
under test.  After each series of measurements over the full range of 
distances, the ambient lighting conditions on the test range were 
changed.  Lighting conditions tested included daylight office conditions 
(i.e., sunlight coming through windows, plus artificial light), normal
fluorescent office lighting, nighttime office lighting (scattered 
fluorescent lights plus some light entering through windows from the 
streetlights outside), and a darkened, windowless conference room.
An optical bandpass filter was used in some tests in an attempt to 
reduce the level of background radiation and determine if detector 
overload was an important factor.  All tests were conducted indoors.

\subsection{Experimental Results}

Results of the survey of devices are shown in Table 
\ref{survey_of_devices_table}.  Of 39 devices tested, 14 showed 
evidence of Class~III optical emanations at the tested bit rate.

\def\ClassOne{& $\bullet$ & $\mbox{ }$ & $\mbox{ }$}
\def\ClassTwo{& $\mbox{ }$ & $\bullet$ & $\mbox{ }$}
\def\ClassThree{& $\mbox{ }$ & $\mbox{ }$ & $\bullet$}

\begin{table}
\centering
\caption{Results of a survey of 39 devices.}
\vspace{.1in}
\label{survey_of_devices_table}
\begin{tabular}{|l|c|c|c|}
\hline
{\it LED Indicator} & {\it Class~I} & {\it Class~II} & {\it Class~III}
\rule{0in}{2.5ex} \\ [0.5ex]
\hline

\multicolumn{4}{|c|}
{\bf Modems and Modem-Like Devices} \rule{0in}{2.5ex} \\ [0.5ex]

\hline

ANP Model 100 short-haul modem, {\tt TD} indicator \ClassThree
\rule{0in}{2.5ex} \\

ANP SDLC card, {\tt TD} indicator \ClassThree \\

CASE/Datatel DCP3080 CSU/DSU, {\tt TD} indicator \ClassThree \\

Hayes Smartmodem OPTIMA 14400, {\tt SD} indicator \ClassThree \\

Hayes Smartmodem OPTIMA 9600, {\tt SD} indicator \ClassThree \\

Motorola Codex 6740 {\tt Hex TP} card, {\tt TD} indicator \ClassThree \\

Motorola Codex 6740 {\tt TP Proc} card, {\tt TD} indicator \ClassThree \\

MultiTech MultiModem V32, {\tt TD} indicator \ClassThree \\

Practical Peripherals PM14400FXMT fax modem, {\tt TX} and & & & \\
{\tt RX} indicators \ClassTwo \\

SimpLAN IS433-S printer sharing device, front panel LEDs \ClassThree \\

Telemet SDR-1000 Satellite Data Receiver, {\tt Data} indicator \ClassThree \\

V.32bis modem simulator, {\tt TD} indicator \ClassThree \\ [0.5ex]

\hline
\multicolumn{4}{|c|}
{\bf LAN Devices} \rule{0in}{2.5ex} \\ [0.5ex]

\hline

3Com TokenLink III Token Ring LAN card, {\tt Link} indicator \ClassOne
\rule{0in}{2.5ex} \\

Cabletron TRXI-24A Token Ring hub, {\tt Activity} indicator \ClassTwo \\

Ethernet NIC, unknown manufacturer, backplane LED \ClassOne \\

Ethernet AUI, unknown manufacturer, {\tt Link} indicator \ClassTwo \\

Ethernet AUI, unknown manufacturer, {\tt Receive} indicator \ClassTwo \\

Ethernet AUI, unknown manufacturer, {\tt Transmit} indicator \ClassTwo \\

Synoptics 2715B Token Ring hub, {\tt Link} indicator \ClassOne \\ [0.5ex]

\hline
\multicolumn{4}{|c|}
{\bf WAN Devices} \rule{0in}{2.5ex} \\ [0.5ex]

\hline

Cisco 4000 IP router, Fast Serial {\tt TD} indicator \ClassThree
\rule{0in}{2.5ex} \\

Cisco 4000 IP router, front panel LED \ClassTwo \\

Cisco 7000 IP router, Fast Serial {\tt TD} indicator \ClassThree \\

Cisco 7000 IP router, front panel LED \ClassTwo \\

Stratacom IPX SDP5080A, {\tt RXD} indicator \ClassTwo \\

Verilink FT1 DSU/CSU, {\tt Pulses} indicator \ClassTwo \\

Westel 3110-30 DS1 Connector, {\tt Pulses} indicator \ClassTwo \\ [0.5ex]

\hline
\multicolumn{4}{|c|}
{\bf Storage Devices} \rule{0in}{2.5ex} \\ [0.5ex]

\hline

2x CD-ROM drive, unknown manufacturer, activity LED \ClassTwo \rule{0in}{2.5ex} \\

Compaq Proliant hot-swappable disk array, activity LED \ClassTwo \\

Compaq Proliant server, floppy drive LED \ClassTwo \\

IBM 4702 controller, 5$\frac{1}{4}$-inch floppy drive LED \ClassTwo \\

IBM 4702 controller, hard disk activity LED \ClassTwo \\

IBM 8580 computer, disk activity indicator \ClassTwo \\

PC, unknown manufacturer, hard disk LED \ClassTwo \\ [0.5ex]

\hline
\multicolumn{4}{|c|}
{\bf Miscellaneous Devices} \rule{0in}{2.5ex} \\ [0.5ex]

\hline

Hewlett-Packard LaserJet 4 laser printer, {\tt Ready} indicator \ClassTwo \rule{0in}{2.5ex} \\

IBM 3745 Front-End Processor, console LEDs \ClassOne \\

IBM 4019 Laser Printer, {\tt Buffer} indicator \ClassTwo \\

IBM 5394-01B Control Unit, {\tt Work Station Active} LED \ClassTwo \\

IBM AS/400 Model 9406, {\tt Processor Activity} indicator \ClassTwo \\

WTI POLLCAT III PBX Data Recorder, {\tt PBX Input A, B} & & & \\
indicators \ClassThree \\ [0.5ex]
\hline
\end{tabular}
\end{table}

\subsubsection{Results of the Survey of Devices}

Dial-up and leased-line modems were found to faithfully broadcast
data transmitted and received by the device.  Only one device of this
type did not 
exhibit Class~III emanations: the Practical Peripherals PM14400FXMT fax 
modem.  The shortest pulse duration measured from this device was 20~ms, 
even at high data rates.

None of the LAN interface cards tested, including 
10~Mbits/s Ethernet and 16 Mbits/s Token Ring adapters, were found to 
broadcast any recognizable data.  Examination of the data sheet for a 
chipset used in fiber optic Ethernet devices reveals a possible reason 
for this finding.  According to \cite{hp_led_data_sheet}, LED drivers 
for {\tt transmit}, {\tt receive}, and {\tt collision} indicators are filtered 
through pulse stretching circuits to make their activity more visible.  
The pulse stretcher extends the {\it on}-time of LED indicators to a minimum 
of several milliseconds.  This makes short pulses easier to see, but 
severely limits the bandwidth of the LED from the perspective of 
compromising optical emanations.  All of the Ethernet and Token Ring 
devices examined showed similar behavior in this regard.

Both of the routers tested (Cisco Series 4000 and 7000 routers equipped 
with Token Ring, Fast Serial and FDDI Interface Processors) were found 
to broadcast Class~III emanations from the Fast Serial
LEDs on their back panels.  Front-panel activity indicators, while 
suggestive of data leakage, typically exhibited a typical minimum pulse 
width on the order of 20~ms, indicating that the front-panel indicators
are merely Class~II.  None of the LAN
devices tested showed any evidence of Class~III emanations from LAN traffic.

Two T1 (1.554~Mbits/s) CSU/DSU (Channel Service Unit/Data Service Unit)
devices were tested.  Neither unit showed evidence of Class~III 
emanations.  Lower-speed CSU/DSU devices, however, on 56~kbits/s leased
circuits, behaved similarly to dial\-up modems.  All showed usable
Class~III emanations in both synchronous and asynchronous operation.

Intelligent serial data switches (i.e., printer sharing devices), a
satellite data receiver, and a PBX call
data recorder behaved similarly to the modems in this test.  Data from 
attached devices showed up in the form of Class~III optical emanations
on the front panels of all these devices.

Mass storage devices such as hard disks and tape transports are usually 
equipped with device activity indicators.  It was hypothesized that the
optical output of 
these LEDs might be related to data transfers to or from the storage 
device.  A variety of PC and minicomputer hard disk drives, floppy 
diskette drives, CD-ROM drives and tape transports were tested.  None 
were found to emit anything other than Class~II optical emanations.
 
Miscellaneous devices tested included the {\tt Processor Activity}
indicator on an IBM AS/400 computer, the {\tt Work Station Active}
indicator on an IBM 5394 terminal controller, and control panel 
indicators on IBM and Hewlett-Packard laser printers.  All of these 
devices were found to be Class~II at most.
 
No significant difference was found between the observability of 5~mm
standard-sized LEDs and the much smaller surface-mount components
used in newer devices.  The absolute brightness levels of these LEDs
are comparable.

\begin{figure}
\centering
\includegraphics[width=\textwidth]{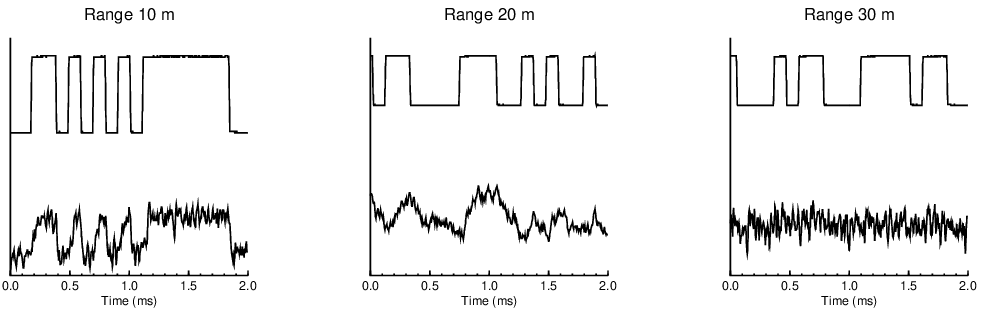}
\caption{Degradation of the optical signal with increasing distance
from the target. Data rate was 9600 bits/s.}
\label{three_captures_in_a_row}
\end{figure}

\subsubsection{Long-Range Testing}

\begin{figure}
\centering
\includegraphics[height=2in]{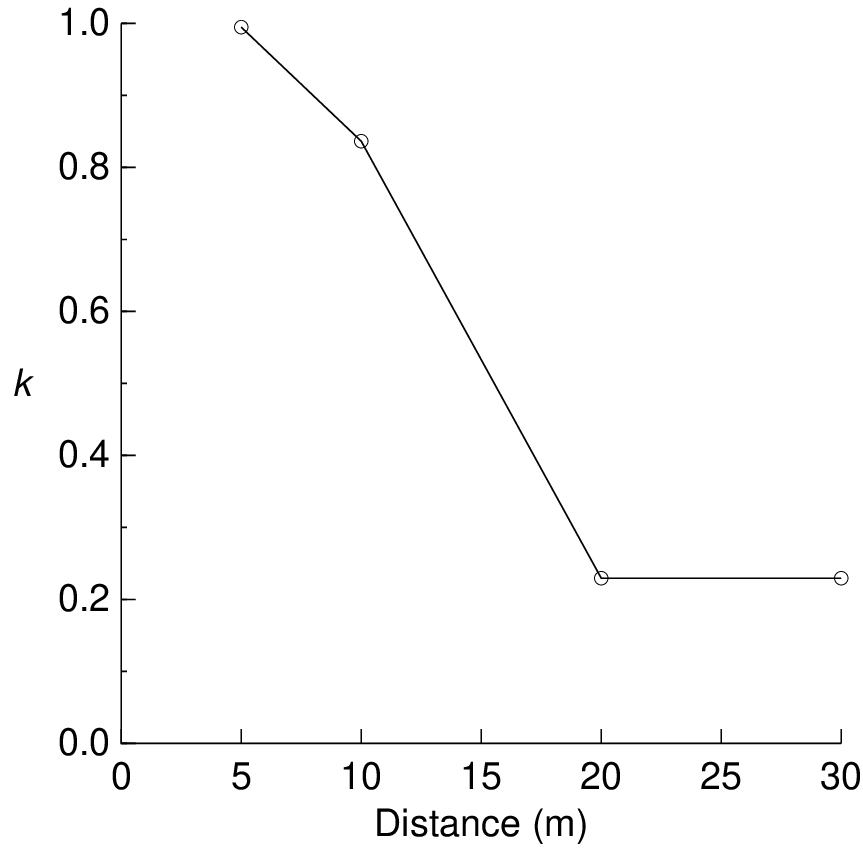}
\caption{Observed correlation $k$ between the original EIA/TIA-232-E
data signal (9600 bits/s) and the received optical signal for
distances of 5~m, 10~m, 20~m, and 30~m.
This is from the data of Figures \ref{example_of_optical_emanations_figure}
and \ref{three_captures_in_a_row}.}
\label{correlation_figure}
\end{figure}

Results of long-range testing are shown in Figure 
\ref{three_captures_in_a_row}.  Note the increasing signal degradation
as the distance was varied from 10~m to 30~m from the detector.  There
is a high correlation evident between the
EIA/TIA-232-E waveform and the received optical signal, as shown in Figure 
\ref{correlation_figure}.  For comparison, the correlation between the 
upper trace of the first part of Figure \ref{three_captures_in_a_row} 
and a random signal of similar amplitude to the optical signal was found to be
$-0.02558$, which is not statistically significant.

No difference was seen at faster bit rates.  Interestingly, several devices
continued to emit a recognizable optical signal at data rates exceeding the
rated capability of the device.  Despite high noise levels in the recorded
waveforms, due apparently to a combination of detector shot noise and thermal
noise in the amplifier, signals were intercepted and properly decoded at a
distance.

\subsubsection{Reverse Engineering of Devices}

It appears that some types of data encryption devices, in particular 
stand\-alone data encryptors and modems with built-in link encryption 
capability, may emit optical signals in unencrypted form.

\begin{figure}
\centering
\includegraphics[width=\textwidth]{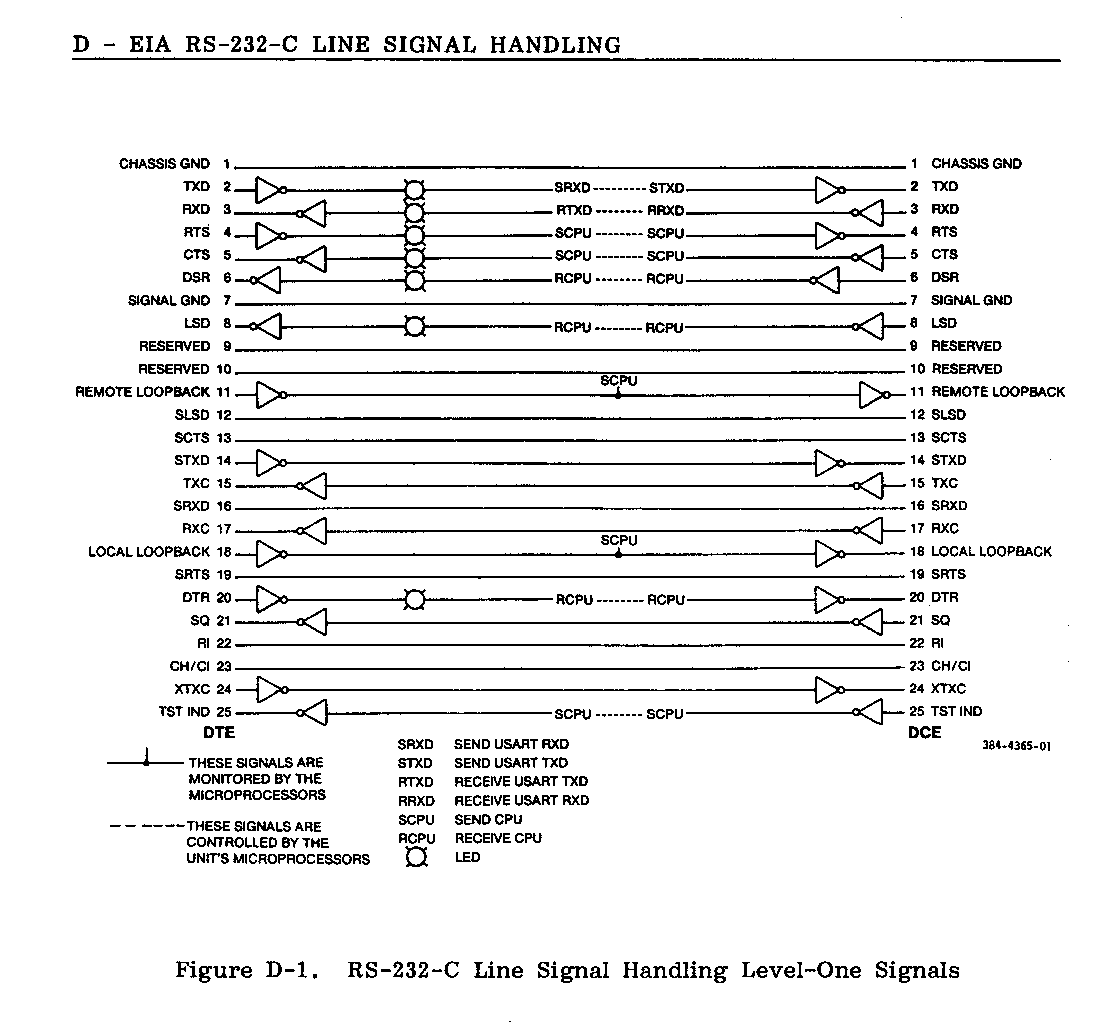}
\caption{Detail from {\it Installation and Operation Manual} 
for the InfoLock 2811-11 DES encryptor.}
\label{infolock_2811_figure}
\end{figure}

Figure \ref{infolock_2811_figure} is a detail taken from the {\it 
Installation and Operation Manual} for the Paradyne InfoLock model
2811-11 DES encryptor.  The InfoLock 2811 is a stand\-alone DES (Data 
Encryption Standard) link encryptor of the type used by financial 
institutions to encrypt data on their wire transfer and ATM (automated
teller machine) networks \cite{Paradyne1985}.

The figure shows a portion of the data path between the DTE connector 
(Data Terminal Equipment---the side of the encryptor that connects to a 
computer) through the encryption function, to the DCE connector (Data 
Communications Equipment---the side that connects to a modem).  The DTE, 
or {\it red} side is unencrypted; the DCE, or {\it black} side is 
encrypted \cite{mil_hdbk_232_a}.  It is clear from this diagram that LED 
indicators on the {\tt TXD} and {\tt RXD} (transmitted and received 
data, respectively) are on the red side of the InfoLock 2811.
This is a serious design 
flaw.  The LEDs will display all of the data passing through the device 
(in either direction) in its unencrypted, or plaintext form.

It is believed that any link encryption device with LED indicators may 
potentially contain this flaw.  Modems with built-in link encryption are 
probably vulnerable as well.  Stand-alone data encryptors like the InfoLock 2811 
will protect downstream equipment on the black side, but are vulnerable
to compromising optical emanations themselves.  The failure mode results
in leakage of cleartext.  The determination of whether or not a 
particular encryption unit is vulnerable will require examination of the 
internals of each device.

\section{Interpretation of Results}

The null hypothesis was disproved.

Class~III emanations were found only in data communication devices, but 
not all data communication devices examined were found to be Class~III.  In particular,
none of the LAN cards tested were found to exhibit Class~III behavior 
(although most of them were Class~II).  No data storage device was found 
to be Class~III.  The design flaw in the InfoLock 2811 encryption device 
is particularly interesting.

Optical background noise from artificial sources proved to be a 
significant problem.  Sources of low-frequency noise (120~Hz and below) 
include incandescent, fluorescent, mercury vapor and sodium vapor lamps; 
high-frequency noise sources include industrial fluorescent lighting and 
compact fluorescent lights.  Sunlight, while dwarfing in brightness the 
artificial sources, contributes only a DC component, and is 
much easier to filter out.  Artificial lighting sources proved to be a 
pervasive and difficult-to-eliminate source of problems,
because they contribute large amounts of amplitude-modulated noise 
containing strong harmonics in precisely the range of interest.  In the 
United States, the standard 60~Hz alternating current electrical 
power supply leads to a characteristic noise component at 120~Hz.  Most 
common data transmission rates are multiples of this frequency, 
complicating recovery of the data.

Digital signal processing techniques can help.  By using a low-pass 
filter to isolate the 120~Hz component of the received optical signal, 
low-frequency noise can be isolated and subtracted from the optical 
signal, yielding a new signal without the 120~Hz component.  Results of 
experiments in this area were very encouraging.   Experiments using 
analog electronic filters were also encouraging.

The limiting factors in long-range interception seem to be the optics 
and the detector--amplifier system.  Both a larger aperture and a narrower 
field of view are required.  It is believed that, out to a range of 
at least several hundred meters, the optical flux available from a
single LED is well within the capability of our detector.  The response
time of a typical LED suggests a practical upper limit on the order of
10 Mbits/s.  Clearly, however, interception of data at longer ranges and
higher speeds is feasible.

\begin{figure}
\centering
\includegraphics[height=2in]{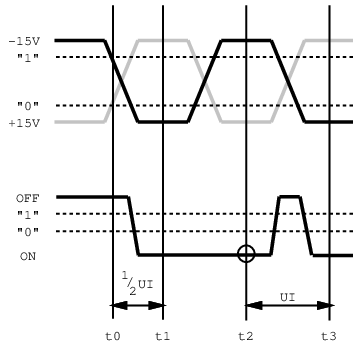}
\caption{Effect of a pulse stretcher between the data source and LED.  
Vertical lines are decision points; the small circle indicates the point
at which an eavesdropper would incorrectly read the optical signal.}
\label{pulse_stretcher_figure}
\end{figure}

\section{Countermeasures}\label{section:countermeasures}

A contributing factor to the threat of optical interception is a
historical tendency to locate computers and data communication equipment 
in environmentally controlled ``glass houses'' which provide no barrier 
to the escape of optical radiation.  Clearly this must now be considered 
a threat.

Examination of lighted windows of high-rise office buildings in the 
evening hours reveals a rich variety of equipment racks with LED 
indicators in view.  Line-of-sight access is surprisingly easy to find.  
Fortunately, optical emanations are easier to contain than RF; opaque 
materials will shield the radiation effectively.
 
Black tape over the LEDs is effective, but inelegant.
The best solution to the problem is a design change.
Status displays could be designed 
to be deactivated when not in use (effectively making them Class~I),
or alternative display technologies could be employed, such as LCD
and displays, which can be made inherently Class~II due to their
relatively slow impulse response.  But many of these other technologies
(such as CRT displays) are more expensive.  LEDs are fast, cheap,
and relatively 
low power indicators that can be read from across a room (a significant 
weakness of liquid crystal displays).  It is preferable to retain these
desirable properties.

A better solution is presented in Figure \ref{pulse_stretcher_figure}.  
The key here is a violation of the worst-case jitter tolerance of the
serial data communication transmission scheme in use \cite{tia_eia_404_b}.
If the minimum {\it on}-time of an LED indicator is 
greater than 1.5 times the unit interval of the current data 
rate\footnote{or alternatively, the slowest data rate expected}, then
an attacker will be unable to recover sufficient information to decode 
the signal.  The effect is to convert a Class~III indicator to Class~II.
The resulting low-pass filter removes a sufficient amount of information from
the optical signal that an attacker cannot recover the original data
from the emanations.  The LED will flicker in response to a random data
signal, and hence will still be useful as a Class~II activity indicator,
but the risk of significant information leakage is reduced.

More conservatively, the minimum
{\it on}-time of the LED could be made to be at least twice the unit interval;
even more conservatively, the minimum {\it off}-time could be similarly
controlled as well.  Most conservatively of all, the minimum {\it on}-time
of the LED should be made to equal the nominal character interval of
the current data rate, or of the slowest data rate expected.  This
will guarantee that an attacker cannot derive any information
from the optical signal other than that a symbol was transmitted.

Even though it appears that at least one device (the PM14400FXMT fax modem)
already incorporates pulse stretching functionality on its status LEDs, it is
believed that this was done to make the display easier to read, not for
reasons of blocking compromising emanations \cite{pulse_stretcher}.

Of course, even in the presence of the aforementioned hardware modification,
a patient attacker might simply time-modulate the asynchronous data stream
in such a way as to effect a covert channel at a rate of
${(\frac{t_\mathrm{UI}}{t_\mathrm{character}})}^{-1}$ bits/s.
It is difficult to completely
eliminate the possibility of covert timing channels in multilevel systems
\cite{pgn_covert_channels_paper}.

\section{Summary and Conclusions}

Modulated optical radiation from LED status indicators appears to be a 
previously unrecognized source of compromising emanations.  This 
vulnerability is exploitable at a considerable distance.  Primarily, data 
communication equipment is affected, although data encryption devices 
also pose a high risk of information leakage, potentially leading to loss of 
plaintext and encryption keys.

A taxonomy of optical emanations was developed according to the amount 
of useful information available to an attacker.  Experiments showed 
that Class~III optical emanations, which should never be permitted, were 
present in 36\% of devices tested, and data could be read from these 
devices at a distance of at least 20~m.  Countermeasures are 
possible that will convert a vulnerable Class~III indicator into the 
safer (but still useful) Class~II variety, by means of inserting a pulse
stretcher into the LED driver circuitry.

\subsection{Conclusions}

Theft of information by interception of optical emanations is necessarily 
limited to one-way---the intruder can only receive information.  
However, login IDs and reusable passwords obtained in this fashion 
could be used to support a conventional attack.  As mentioned before, 
parity checking, CRC values, and other error detection and correction features 
embedded in the data stream are available to the eavesdropper too, and can 
be of great benefit in helping to overcome the effects of a low-quality 
optical signal.

Ironically, it may be the simplest devices---low-speed, 
obsolete, insignificant parts of a network---that provide a gateway 
for intruders.  In our experiments, it was low-speed modems, 
routers, line drivers, data loggers, and a printer sharing device that
were found to be the most enthusiastic broadcasters of data. 
Class~III optical emanations have been observed in the wild from
devices as diverse as TTY-equipped payphones in airports and the
digital control box of a player piano.  Like the Purloined 
Letter, they hide in plain sight: a tangle of 
remote office connections in the corner, a modem sitting next to a PC by 
the window, or a call-accounting system on the PBX.

\subsection{Summary of Contributions}

\begin{itemize}

\item The existence of compromising optical emanations was proved.

\item Successful exploitation, under realistic conditions, was 
demonstrated.

\item A taxonomy of compromising optical emanations was developed.

\item Some possible countermeasures were presented.

\end{itemize}

\section{Directions for Future Research}

Much work remains to be done.  While we have shown that it is possible
to intercept data at realistic data rates out to a few tens of meters,
the maximum distance at which this can be accomplished remains unknown.
Improved signal detection techniques, optics, and detectors would go
a long way toward quantifying the effective limits on distance and bit
error rate.

Other possible areas of investigation include the exploitation of Class~II
devices (especially disk drive and LAN card activity indicators) by
covert channels; methods for dealing with extremely low-level optical
emanations; exploitation of non--line-of-sight, or diffuse, emanations;
several interesting aspects of
fiber optics, including dark fiber; and the opportunities afforded by
stimulated emanations (Appendix \ref{appendix:covert_channel}).

\subsection{Low-Level Optical Emanations}

While no evidence was found of Class~III emanations from data storage 
devices, more investigation is needed to verify that disk and tape drive 
activity indicators experience no second-level effects due to (for instance)
insufficient power supply regulation.  And the wide variety and 
distribution of LAN cards suggests the possibility that at least one 
might show more than just Class~II activity on the LEDs\footnote{LAN 
cards on PCs are particularly interesting, given that the LEDs are on 
the back panel; when the computer is conventionally oriented on a desk
by a window, the LEDs are clearly visible from outside.}.

Other possible sources of compromising optical emanations include 
leakage from improperly terminated fiber optics or unconnected fiber 
optic ports.  Alternative forms of attack are possible, including active 
attacks via optically emitting bugs operating outside of the visible 
spectrum that would be missed by conventional (RF) countersurveillance 
scanners.  A passive collection system could operate over dark fiber;
an accidental passive fiber optic tap would result if the end of a fiber
strand were exposed to optical emanations from other devices in the room.

\subsection{The Possibility of Non--Line-of-Sight Interception}

\begin{figure}
\centering
\includegraphics[height=2in]{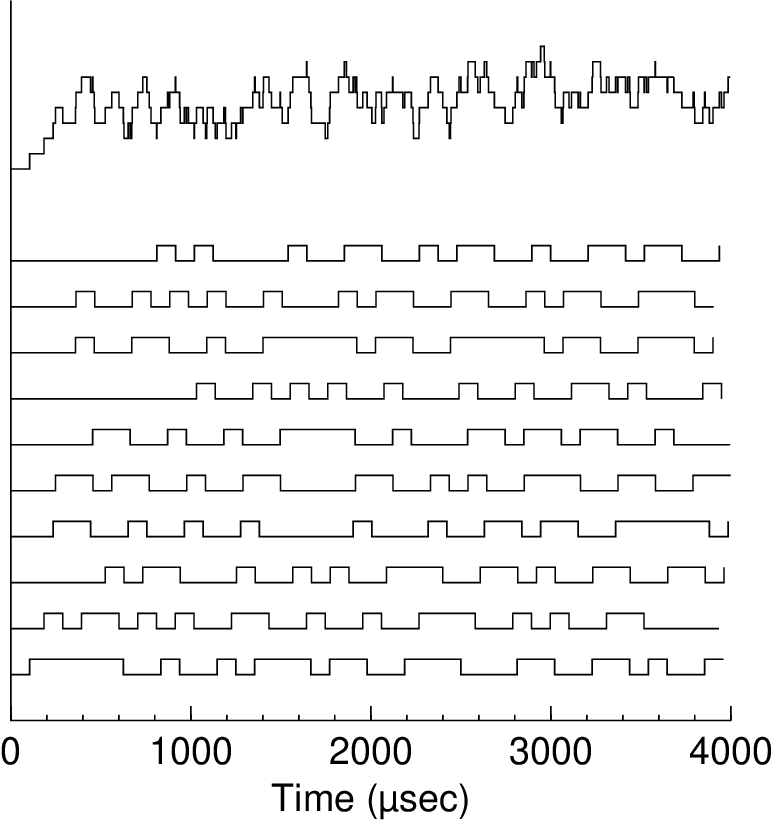}
\caption{Optical sum of ten random but correctly formatted data
signals.  {\bf SIMULATION}}
\label{diffuse_emanations_figure}
\end{figure}

Still unresolved is the question of whether diffuse emanations from 
multiple commingled or non--line-of-sight sources can be profitably
unraveled.  Optical signals sum linearly; the optical sum of $n$ linearly
independent
sources results in an extremely complicated signal
(Figure \ref{diffuse_emanations_figure}).  A room full of
LED status indicators, even if individual sources
are not directly observable, nevertheless can be seen to fill an entire area with
a diffuse red glow.  Light leakage around a door, or a passive fiber optic 
tap\footnote{A passive fiber optic tap might consist of as little as an
unused strand of fiber, terminated at a patch panel inside the room but
reserved for future use.  Consequently, unused fiber optic ports should be
capped when not in use.}, might provide an adversary with enough optical
flux to begin to analyze it.

\begin{table}
\centering
\caption{Decoding of diffuse emanations from state transitions. \bf{SIMULATION}}
\vspace{.1in}
\label{decoding_diffuse_emanations_table}
\begin{tabular}{|c|c|c|c|c|}
\hline
$t_\mathrm{event} \mathrm{(\mu s)}$ & {\it Transition} & {\it Interpretation} & \
$t_\mathrm{next-event} \mathrm{(\mu s)}$ & {\it What can be deduced?}
\rule{0in}{2.5ex} \\ [0.5ex]
\hline
$104.1\overline{6}$ & $\uparrow$ & $\phi_1$ start bit & $208.\overline{33}$ & \rule{0in}{2.5ex} \\
$184.1\overline{6}$ & $\uparrow$ & $\phi_2$ start bit & $288.\overline{33}$ & At least two signals exist.\\
$208.\overline{33}$ & none & $\phi_1$ $\mathrm{data}_0 = 1$ & $312.50$ & \ \\
$235.1\overline{6}$ & $\uparrow$ & $\phi_3$ start bit & $339.\overline{33}$ & \ldots three signals \ldots \\
$248.1\overline{6}$ & $\uparrow$ & $\phi_4$ start bit & $352.\overline{33}$ & \ldots four signals \ldots \\
$288.\overline{33}$ & $\downarrow$ & $\phi_2$ $\mathrm{data}_0 = 0$ & $392.50$ & \ \\
$312.50$ & none & $\phi_1$ $\mathrm{data}_1 = 1$ & $416.\overline{66}$ & \ \\
$339.\overline{33}$ & none & $\phi_3$ $\mathrm{data}_0 = 1$ & $443.50$ & \ \\
$352.\overline{33}$ & none & $\phi_4$ $\mathrm{data}_0 = 1$ & $456.50$ & \ \\
$359.1\overline{6}$ & $\uparrow$ & $\phi_5$ start bit & $463.\overline{33}$ & \ldots five signals \ldots \\
$362.1\overline{6}$ & $\uparrow$ & $\phi_6$ start bit & $466.\overline{33}$ & \ldots six signals \ldots \\
$392.50$ & $\uparrow$ & $\phi_2$ $\mathrm{data}_1 = 1$ & $496.\overline{66}$ & \ \\
$416.\overline{66}$ & none & $\phi_1$ $\mathrm{data}_2 = 1$ & $520.8\overline{3}$ & \ \\
$443.50$ & $\downarrow$ & $\phi_3$ $\mathrm{data}_1 =  0$ & $547.\overline{66}$ & \ \\
\hline
\end{tabular}
\end{table}

\subsubsection{Experiments with Non--Line-of-Sight Access}

Sometimes things that are impossible in theory turn out to be feasible
in practice.  While it is true that in the general case of $n$ random square
wave signals---whose amplitude, pulse width, and pulse repetition rate are
unknown---that a unique decomposition may not exist, in the real world, however,
data signals are {\it not} random.  EIA/TIA-232-E in particular is full
of known values: start symbols, stop symbols, the number of data bits following
a start symbol, and the guaranteed minimum and maximum duration of all symbols
(the unit interval).  Because the signals are not entirely random, but contain
a small number of known values at certain fixed locations, it becomes possible
to identify individual components ($\phi_1$ through $\phi_n$) with high
probability.  The left-hand side of Table \ref{decoding_diffuse_emanations_table}
gives the timing and direction of state transitions during the first few hundred
$\mathrm{\mu s}$ of the simulation shown in Figure \ref{diffuse_emanations_figure}.
The algorithm works by scanning the received optical waveform from left to
right until the first positive-going transition is found.  The optical signal
should be sampled at a rate at least three to five times the reciprocal of the
smallest time difference between successive level transitions \cite{McCarthy}.
Once a candidate transition is identified in the aggregate optical sum, any further
activity on that particular component ($\phi_i$) can be ruled out for at least
one unit interval.  Any transitions seen in the meantime must be the result of
another, heretofore unknown signal ($\phi_{i+1}$).  By an iterative process of
elimination, each individual signal in turn is teased out of the jumble.

\begin{figure}
\centering
\includegraphics[height=2in]{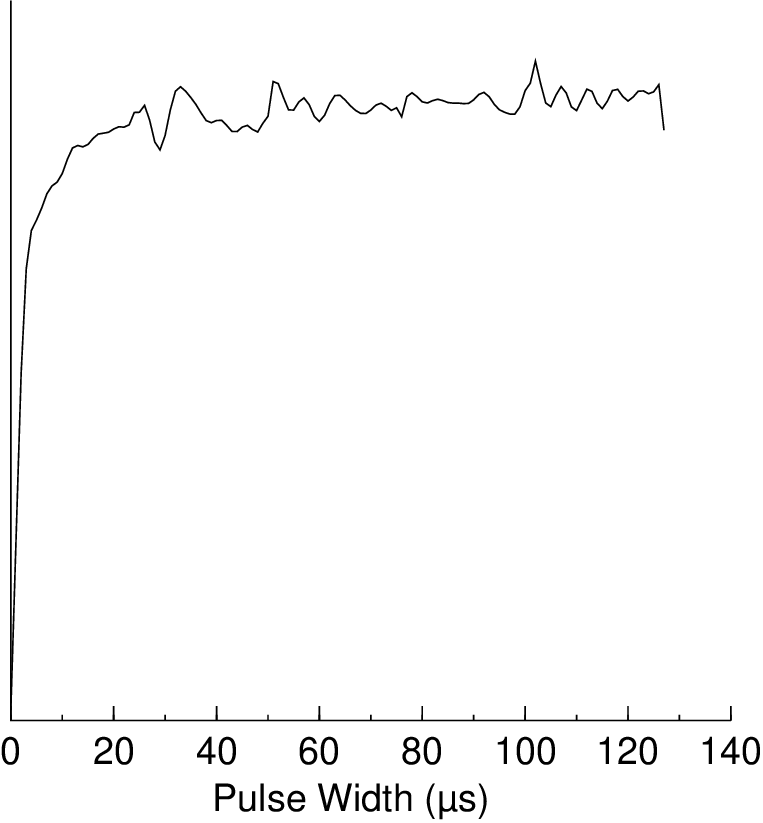}
\caption{Fourier spectrum (real part) of the interval between transitions
in the optical sum in Figure
\ref{diffuse_emanations_figure}.  The peak in the curve, at approximately
$104 \ \mathrm{\mu s}$, corresponds to the most likely unit interval.}
\label{fourier_spectrum_figure}
\end{figure}

The unit interval is not difficult to guess from the Fourier spectrum of the
times of transitions in the received signal
(Figure \ref{fourier_spectrum_figure}).  The peak in the curve, at
$104 \ \mathrm{\mu s}$, corresponds to the most likely unit interval.  In
any case, the range of possible data formats is small enough simply
to try all of the various possibilities until one of them yields intelligible data.
Even an ambiguous solution might be of some value to an attacker, if the result
were a data stream having some non-zero, but not catastrophic, bit error
rate\footnote{For example, if the most-significant bit of an 8-bit data word
is not always 0, the data stream is not ASCII.  Similarly, there are many
disallowed values in EBCDIC that could be used to rule out this encoding as well.}.
As long as the individual signal components are not precisely aligned in
time---which leads to ambiguous solutions---the analysis appears to be tractable.
More work is clearly needed, on real signals, as a follow-on to the unrealistically
low-noise simulation presented here.

\bibliographystyle{acmtrans}
\bibliography{optical_tempest}

\appendix

\section{Stimulated Emanations}\label{appendix:covert_channel}

Not all sources of compromising optical emanations are naturally 
occurring.  We describe two implementations of a Trojan horse that 
manipulates the LEDs on a standard keyboard to implement a
high-bandwidth covert channel \cite{wray}.  This is an example of an
{\it active attack}, mounted by an adversary against a device
that is not normally vulnerable to compromising optical emanations.

\subsection{The Keyboard Considered as an Output Device}

Ever since the standardization of computer keyboards to the IBM layout, 
most computer keyboards have three LED indicators, for Caps Lock, 
Num Lock, and Scroll Lock, respectively.  Interestingly, these LEDs are
not directly connected to their associated keys---the lights, in fact,
are software controlled.

The PC keyboard is an intelligent device that communicates with the host 
computer over a bi-directional, synchronously clocked serial interface at 
approximately 10\,000~bits/s\footnote{The exact speed was found 
to vary among different manufacturers.}.

The capacity of the keyboard interface channel far exceeds the 
requirements of even the fastest typist.  So long as the amount of data 
sent to the keyboard is limited, and does not interfere with processing 
of keystrokes, the excess bandwidth can be profitably employed by an 
attacker.

A {\it covert channel} is a means of extracting data from a computer 
system in violation of the system security policy
\cite{lampson,ncsc_covert_channels_book}.  A 
{\it high-bandwidth} covert channel is considered to be one capable of 
transmitting data faster than 100~bits/s
\cite{common_criteria,orange_book}.  The covert channel described here
has been demonstrated to work at speeds from 150 to 10\,000~bits/s.

\subsection{Related Work}

The fact that keyboard LEDs can be manipulated has been known for a long
time.  Some operating systems provide
the capability to control the keyboard indicators from a shell script; if 
not, then it is a relatively simple matter to program directly to the 
keyboard interface \cite{van_gilluwe}.

A more recent paper describes another possible method for remotely monitoring
the electrical signals inside a PC keyboard, together with some
countermeasures. \cite{nato_soft_tempest}.
The only other published description of a covert channel employing keyboard
LEDs appears in a work of fiction \cite{cryptonomicon}, in which a character
employs a similar 
technique to extract a small amount of critical information from his 
computer despite being under continuous surveillance.

\subsection{A Covert Channel in Software }

A successful covert channel running at up to 450~bits/s was
demonstrated on the IBM PC/AT, several different Compaq ProLineas, and
the Sun Microsystems SPARCstation~20 and 
Ultra~1 workstations.  The attack was successful under MS-DOS, Microsoft 
Windows 3.1, Windows 95, and Windows 98, Windows NT 3.5 and 4.0, and Sun
Microsystems Solaris 2.5, 2.5.1, Solaris 7, and Trusted Solaris 2.5
and 2.5.1.
A handful of machines could not be made to work, among them a Compaq
LTE Lite 486/25E notebook.

We found that activity on a single keyboard LED at 150~bits/s was not 
particularly noticeable during interactive use.  Employing more than one 
LED at a time increases the probability of discovery but offers some
compelling
advantages.  If all three LEDs are modulated identically, the optical 
output of the transmitter is tripled, greatly increasing the useful 
range.  Alternatively, two or even three bits could be transmitted in 
parallel, increasing the bandwidth of the covert channel to 
approximately 450~bits/s.  Experiments were run with (1) asynchronous parallel 
transmission using three LEDs, (2) synchronous serial transmission using
single and biphase clocking, and (3) differential Manchester 
encoding.  The latter yielded high reliability at the receiving end, but 
with all the activity on three LEDs, it was noticeable to the 
operator that something strange was going on.

Appendix \ref{appendix:code} contains example code implementing the covert
channel under Solaris version 2.$x$
\footnote{The authors demonstrated this technique privately in 1996.}.

\subsection{Attacking the Hardware}

\begin{figure}
\centering
\includegraphics[height=2in]{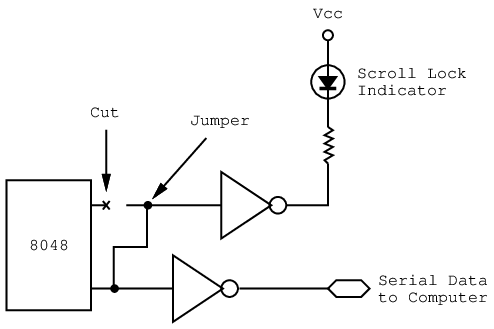}
\caption{Modifications to the IBM PC/AT keyboard.}
\label{keyboard_cross_connect_figure}
\end{figure}

Even better results can be obtained through modification of the keyboard 
hardware.  Depending on the details of a particular keyboard, the 
modifications may be as simple as moving a single wire.  An IBM PC/AT 
keyboard was modified as an experiment.  The Scroll Lock LED was
cross-connected to the {\tt keyboard data} signal, as shown in Figure 
\ref{keyboard_cross_connect_figure}.  It was necessary to invert the {\tt
keyboard data} signal so that the LED would remain dark when the covert 
channel was idle.  Fortunately, IBM provided a ready-made 
solution in the form of an unused gate on one of the chips.  The optical 
output of the LED is now modulated directly by the 10\,000~bits/s
serial data stream in the keyboard cable.  The Scroll Lock LED can be 
seen to flicker momentarily with keyboard activity, but the effect is
not very noticeable.  No software is required.

Normal operation of the Scroll Lock LED is prevented, but the Scroll Lock
function is not used very often.  By a fortuitous coincidence, the normal 
behavior of the keyboard LEDs during the power-on self test (POST)
function is unaffected; the functionality of the Scroll Lock key itself
is also unchanged (except of course that the LED does not appear to work
anymore.)

\begin{figure}
\centering
\includegraphics[height=2in]{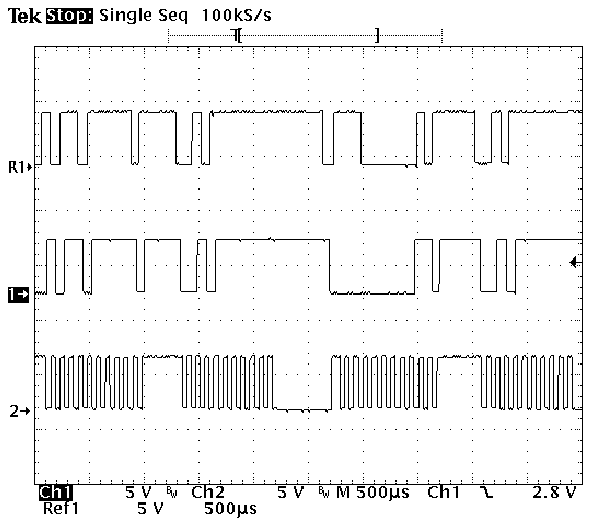}
\caption{Optical signal (top) obtained from a keyboard with the 
modifications of Figure \ref{keyboard_cross_connect_figure}.}
\label{keyboard_intercept_figure}
\end{figure}

Figure \ref{keyboard_intercept_figure} shows the optical waveform 
obtained from a keyboard with the modifications of Figure 
\ref{keyboard_cross_connect_figure}.  The upper trace of
Figure \ref{keyboard_intercept_figure} shows the
intercepted optical signal; the lower two traces are the electrical
signals on the keyboard data interface and the keyboard data clock.
The bandwidth of the resulting 
covert channel is greater than that of a software-only attack, but the 
information is in the form of keyboard scan codes, not ASCII.  It 
requires a bit of translation on the receiving end, but also yields more 
information.  Since accurate timing of both key-down and key-up events 
are reported, this method may provide enough information to compromise 
identity verification systems based on typing characteristics
\cite{umphress_and_williams} or the
generation of cryptographic keys \cite{pgp}.

\subsubsection{Improving the Bandwidth of the Covert Channel}

It is not difficult to imagine how a small investment in additional 
hardware could vastly improve the chances of a successful attack.  An 
infrared (IR) LED chip could be co-encapsulated with a visible LED in 
the same package.  If the two LEDs were connected back-to-back 
internally, only two leads would be required, and the Trojan LED would 
be indistinguishable from a standard component except under high 
magnification.  Modification to the keyboard controller circuitry would 
be required to utilize the IR capability; as long as this is being done 
anyway, the following ``improvements'' might be made to the controller's 
firmware at the same time:

\begin{itemize}

\item Increasing the drive current to the IR emitter for correspondingly 
increased range

\item Use of more sophisticated channel encoding to reduce transmission 
errors and support higher speeds

\item A timer and buffer memory to allow for a delay in sending until 
the keyboard has been idle for a while

\item Encryption and compression of the covert channel data

\item Sender identification, to support multiple units in a single location

\item Pattern matching capability, to look for specific information in 
the keyboard data stream

\item Preserving the normal functionality of the visible LED indicator.

\end{itemize}

All but the first of these have been successfully demonstrated in
software.  Given that access to the hardware or surreptitious replacement
would be necessary in order to emplace a hardware Trojan horse, concurrent 
implementation of all the above features would seem to pose little trouble.  
Modifications to firmware would be nearly undetectable, barring a close 
examination of the microcontroller object code.

\subsection{Conclusions}

This vulnerability potentially affects hundreds of millions of devices.
It might be argued that keyboard LEDs lack sufficient brightness to be 
successfully exploited from a long distance.  However, the authors once 
encountered a Compaq PC whose keyboard LEDs were bright enough to throw 
shadows on the ceiling.  When tested, this keyboard was able to handle 
450~bits/s communication on all three LEDs simultaneously without 
noticeably affecting response time.  The software presented in
Appendix~\ref{appendix:code} is small enough to be included in a computer
virus, as described in \cite{information_hiding}.

\subsection{Summary}

It has been shown that it is possible to cause the emission of 
compromising optical emanations in devices not normally vulnerable, 
by taking advantage of software-controlled LED indicators.
The covert channel thereby 
created has a bandwidth of several hundred bits/second at least, and is 
compatible with standard techniques for exploiting compromising optical 
emanations described in the previous paper.

\section{Sending Data Through the Keyboard}\label{appendix:code}

The following C program implements the software version of the covert
channel under 
Solaris version 2.$x$.  It transmits ASCII data by modulating the Caps Lock LED
with serial data at 50~bits/s.  A similar program, written in Intel
x86 assembly language and incorporating additional 
functionality\footnote{The program installed itself as an interrupt 
handler and hooked the keyboard interrupt.  It copied all keyboard 
activity while waiting for the keyboard to become idle.  After four 
hours of no keyboard activity, the contents of the buffer were
transmitted.  If any keyboard activity was detected while the program
was busy transmitting, it immediately stopped sending, restored the
state of the keyboard LEDs, and resumed waiting.}, required less than
1500 bytes of memory.

\small{
\begin{verbatim}
/*
// sl.c -- a covert channel using the Caps Lock LED.
//
// For Solaris 2.x on SPARC; compile with ${CC} sl.c -lposix4
*/

#include <fcntl.h>
#include <stdio.h>
#include <stdlib.h>
#include <sys/kbio.h>
#include <sys/kbd.h>
#include <time.h>
#include <unistd.h>

#define SPEED 50 /* data transmission speed (bits per second) */

void set_led (int fd, char *data);
void time_led (int fd, char *data);
void perror_exit (char *function_name);

/* set up a 20 millisecond intersymbol delay */

struct timespec min, max = { 0, 1000000000 / SPEED };

int
main (void)
{
    char message[] = "My credit card number is 1234 5678 910 1112.";
    char restore_data;
    char *p = &message[0];
    int  fd;

    /* open the keyboard device */
    if ((fd = open ("/dev/kbd", O_RDONLY)) < 0)
        perror_exit ("open");

    /* save the state of the keyboard LEDs */
    if (ioctl (fd, KIOCGLED, &restore_data) < 0)
        perror_exit ("ioctl");

    while (*p) {
        char data = LED_CAPS_LOCK;
        int  i;

        /* start bit is a "1" */
        time_led (fd, &data);

        /* send 8 bits, least significant first */
        for (i = 0; i < 8; i++) {
            data = *p >> i & 1 ? LED_CAPS_LOCK : 0;
            time_led (fd, &data);
        }

        /* stop bit is a "0" */
        data = 0;
        time_led (fd, &data);

        /* next character of message */
        p++;
    }
    /* restore state of the keyboard LEDs */
    set_led (fd, &restore_data);

    return (close (fd));
}

/* turn keyboard LEDs on or off */

void
set_led (int fd, char *data)
{
    if (ioctl (fd, KIOCSLED, data) < 0)
        perror_exit ("ioctl");
}

/* transmit one bit */

void
time_led (int fd, char *data)
{
    set_led (fd, data);
    nanosleep (&min, &max);
}

/* display an error message and quit */

void
perror_exit (char *function_name)
{
    perror (function_name);
    exit (1);
}
\end{verbatim}
}

\begin{acks}
The authors wish to thank the anonymous reviewers; their careful
reading and insightful comments helped catch a number of errors
that otherwise would have crept into publication.  Annie Cruz of
Washington Mutual Bank, and Eduard Telders of PEMCO Financial
Services, also provided valuable assistance and encouragement with
the preparation of this paper.
\end{acks}

\received{Received April 2001; revised February 2002 and March 2002;
accepted March 2002}

\end{document}